# Tunable RF and microwave photonic sideband generator based on cascaded 49GHz and 200GHz integrated ring resonators

Xingyuan Xu,[1,8] Jiayang Wu,[1,8] Linnan Jia,[1] Mengxi Tan,[1] Thach G. Nguyen,[2] Sai T. Chu,[3] Brent E. Little,[4] Roberto Morandotti,[5,6,7] Arnan Mitchell,[2] and David J. Moss[1]

[1]Centre for Micro-Photonics, Swinburne University of Technology, Hawthorn, VIC 3122, Australia
[2]ARC Centre of Excellence for Ultrahigh-bandwidth Devices for Optical Systems (CUDOS), RMIT University, Melbourne, VIC 3001, Australia
[3]Department of Physics and Material Science, City University of Hong Kong, Hong Kong, China.
[4] State Key Laboratory of Transient Optics and Photonics, Xi'an Institute of Optics and Precision Mechanics, Chinese Academy of Science, Xi'an, China.
[5]INRS-Énergie, Matériaux et Télécommunications, 1650 Boulevard Lionel-Boulet, Varennes, Québec, J3X 1S2, Canada.
[6]National Research Uni of Information Technologies, Mechanics and Optics, St. Petersburg, Russia.
[7]Institute of Fundamental and Frontier Sciences, University of Electronic Science and Technology of China, Chengdu 610054, China.
[8]These authors contribute equally to this paper.
E-mail: dmoss@swin.edu.au

**Abstract**

We demonstrate a continuously RF tunable orthogonally polarized optical single sideband (OP-OSSB) generator based on dual cascaded micro-ring resonators. By splitting the input double sideband signal into an orthogonally polarized carrier and lower sideband via TE- and TM-polarized MRRs, an OP-OSSB signal is generated. A large tuning range of the optical carrier to sideband ratio of up to 57.3 dB is achieved by adjusting the polarization angle of the input light. The operation RF frequency of the OP-OSSB generator can be continuously tuned with a 21.4 GHz range via independent thermal control of the two MRRs. Our device represents a competitive approach towards OP-OSSB generation with wideband tunable RF operation, and is promising for photonic RF signal transmission and processing in radar and communication systems.

## 1. Introduction

Microwave photonics has attracted great interest in a wide range of applications in radar and communications systems [1–4] due to its numerous intrinsic advantages such as broad RF operation bandwidth, low loss, and strong immunity to electromagnetic interference [5-8]. As one of the key technologies of microwave photonic systems, modulation formats have a significant impact on the overall system performance [9-11]. Offering advantages such as overcoming dispersion-induced distortion, enhanced spectral efficiency, and the ability to separately manipulate the optical carrier and sidebands via polarization-sensitive optical components, orthogonally polarized optical single sideband (OP-OSSB) modulation has been widely exploited in applications ranging from antenna beamforming to microwave photonic signal processing [12-16].

Many approaches have been demonstrated to realize OP-OSSB generation, including using specially designed modulators such as acoustic optical modulators [17], Sagnac-loop-based modulators [18], dual-polarization quadrature phase shift keying modulators [19], and polarization modulators [20]. In other approaches, differential group delay elements were used to cross-polarize the optical carrier and the sideband by exploiting birefringence [15,16]. In addition, fiber-based stimulated Brillouin scattering has been used to control the polarization of optical signals to achieve OP-OSSB generation [21,22]. However, the above approaches face limitations of one form or another. On the one hand, RF couplers can introduce an electrical bandwidth bottleneck for the entire system, while fiber-based methods face limitations

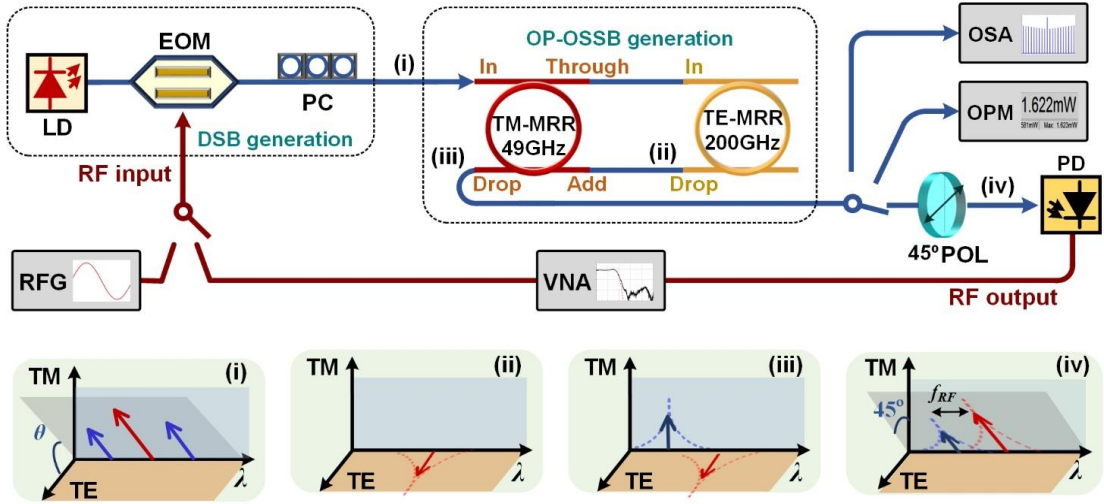

Fig. 1. Schematic of the proposed orthogonally polarized optical single sideband (OP-OSSB) generator. LD: laser diode. EOM: electro-optical modulator. PC: polarization controller. DSB: double sideband. OSA: optical spectrum analyzer. OPM: optical power meter. 45º POL: optical polarizer with the polarization direction having an angle of 45º to the TM axis. PD: photodetector. VNA: vector network analyzer. RFG: RF generator.

in terms of footprint and stability, important for moving the technology beyond the laboratory.

Integrated microwave photonics is a competitive solution to address these challenges [23], offering a reduced footprint and higher reliability [24]. Recently, we reported an OP-OSSB generator that operated at fixed RF frequencies based on a micro-ring resonator (MRR) that supported both TE- and TM-polarizations [25]. Here, we report a continuously tunable (in RF frequency) OP-OSSB generator based on dual cascaded integrated MRRs. The operation RF frequency of the OP-OSSB generator, determined by the spectral interval between the TE and TM resonances, can be dynamically tuned via separate thermo-optical control of the two MRRs, resulting in operation over a wide RF tuning range. Moreover, by controlling the polarization angle of the input light, a large dynamic tuning range in the optical carrier-to-sideband ratio (OCSR) of up to 57.3 dB is demonstrated.

## 2. Operation principle

Figure 1 shows a schematic of the wideband tunable OP-OSSB generator. Continuous-wave light from a tunable laser source is intensity-modulated by an RF signal to generate a double sideband signal with a polarization angle $\theta$ to the TE-axis (Fig. 1(i)), and then fed into the two cascaded MRRs that support TE- and TM-polarization modes. When the wavelength of the optical carrier and the input RF frequency match with the orthogonally polarized resonances of the two MRRs, the optical carrier and one sideband of the double sideband signal can be dropped (Fig. 1(ii)). After that, the dropped optical carrier and sideband are combined together by connecting the drop-port of the second (200GHz, TE polarization) MRR to the add-port of the first (49GHz, TM) MRR (Fig. 1(iii)), thus achieving OP-OSSB modulation.

To analyze the polarization states of our device, we use the Jones matrix formalism. While other methods exist, such as the Poincaré sphere and Stokes parameters [26], we use the Jones matrix for simplicity since the polarization eigenmodes of the MRRs serve as a natural basis. The transmission of the dual MRRs can be written as

$$R = \begin{bmatrix} D_{TE} & 0 \\ 0 & D_{TM} \end{bmatrix} \quad (0.1)$$

where $D_{TE}$ and $D_{TM}$ are the drop-port transfer functions of the 49GHz (TE) MRR and 200GHz (TM) MRR given by

$$D_{TE} = \frac{-k_{TE}^2 \sqrt{a_{TE}} e^{i\phi_{TE}/2}}{1 - t_{TE}^2 a_{TE} e^{i\phi_{TE}}} \quad (0.2)$$

$$D_{TM} = \frac{-k_{TM}^2 \sqrt{a_{TM}} e^{i\phi_{TM}/2}}{1 - t_{TM}^2 a_{TM} e^{i\phi_{TM}}} \quad (0.3)$$

where $t_{TE}$, $t_{TM}$, $k_{TE}$ and $k_{TM}$ are the field transmission and cross-coupling coefficients between the bus waveguide and the ring ($t^2 + k^2 = 1$ for lossless coupling), $a_{TE}$ and $a_{TM}$ represent the round-trip transmission factors, $\phi_{TE} = 2\pi L_{TE} \times n_{\text{eff\_TE}}/\lambda$ and $\phi_{TM} = 2\pi L_{TM} \times n_{\text{eff\_TM}}/\lambda$ are the single-pass phase shifts of the TE-MRR and TM-MRR, respectively, with $L_{TE}$ and $L_{TM}$ denoting the round-trip length, $n_{\text{eff\_TE}}$ and $n_{\text{eff\_TM}}$ denoting the effective indices, and $\lambda$ denoting the wavelength.

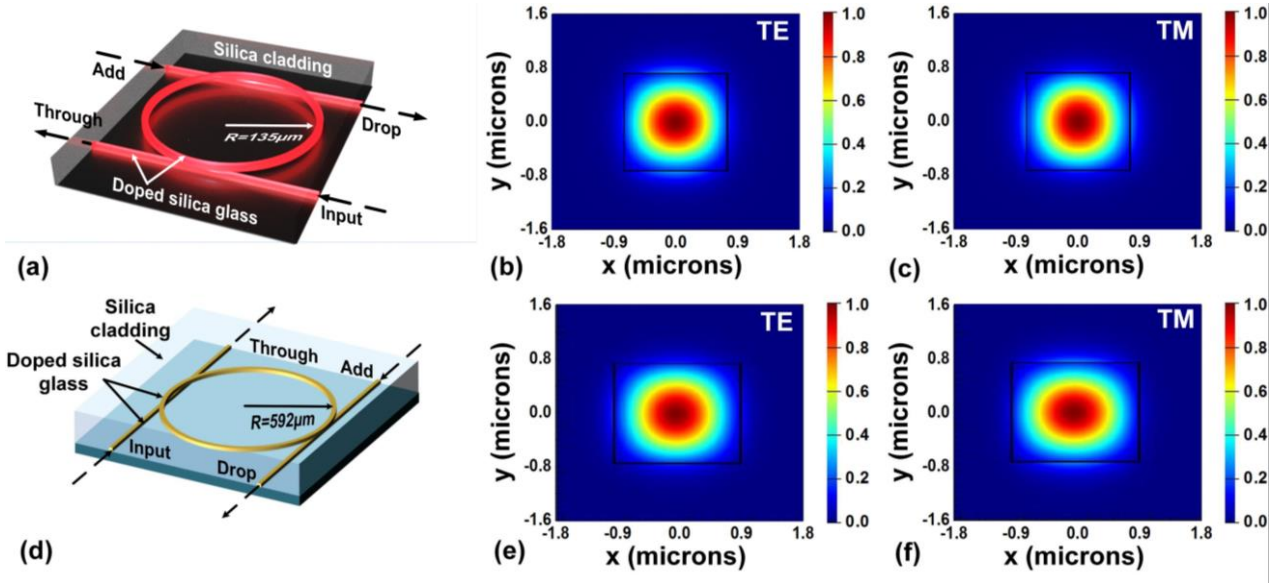

Fig. 2. (a) Schematic illustration of the 200GHz FSR MRR, (b) TE and (c) TM mode profiles of the 200GHz MRR. (d) Schematic illustration of the 49GHz FSR MRR, (e) TE and (f) TM mode profiles of the 49GHz FSR MRR.

For a general optical field input $E_0 \begin{bmatrix} \cos\theta \\ \sin\theta \end{bmatrix}$, the output field of the dual MRRs is

$$E_{out} = RE_0 \begin{bmatrix} \cos\theta \\ \sin\theta \end{bmatrix} = E_0 \begin{bmatrix} D_{TE} \cdot \cos\theta \\ D_{TM} \cdot \sin\theta \end{bmatrix} \qquad (0.4)$$

As reflected by the above equation, the optical power of the spectral components dropped by the 49GHz (TE) MRR and 200GHz (TM) MRR are proportional to $\cos^2\theta$ and $\sin^2\theta$, respectively. Thus, the OCSR (with the 49GHz MRR for the carrier and the 200GHz MRR for the sideband) is given by

$$\text{OCSR}(\theta) \propto \cot^2\theta \qquad (0.5)$$

which can be continuously tuned by adjusting $\theta$. Since $\cot^2\theta$ can infinitely approach 1 or 0 as $\theta$ approaches 0 or $\pi/2$, a large tuning range of the OCSR can be produced. Moreover, the generated OP-OSSB signal can be converted back into the RF domain by passing it through an optical polarizer (Fig. 1(iv)). The operation RF frequency of the proposed OP-OSSB generator is determined by the spectral interval between adjacent resonances of the 49GHz and 200GHz MRR. Thus, with separate thermal controls of the MRRs, tunable OP-OSSB generation over a wide RF tuning range can be realized.

## 3. Experimental results

We fabricated 49GHz FSR and 200GHz FSR integrated MRRs that each support both TE- and TM- polarization modes, utilizing the TM-polarized resonances of the first MRR (49GHz), and the TE-polarized resonances of the second (200GHz) MRR. Figures 2(a, d) show a schematic of the two MRRs used in our experiment, which were both fabricated on a high-index doped silica glass platform using CMOS compatible fabrication processes [27-29]. First, high-index (n = ~1.70 at 1550 nm) doped silica glass films were deposited using standard plasma enhanced chemical vapour deposition, then patterned using deep UV photo-lithographically and etched via reactive ion etching to form waveguides with exceptionally low surface roughness [30-32]. Finally, silica glass (n = ~1.44 at 1550 nm) was deposited as an upper cladding. The waveguides were designed to feature a nearly symmetric cross-section (1.5 μm × 2 μm for the 49GHz ring resonator [33] and 1.45 μm × 1.5 μm for the 200GHz ring resonator [29]), enabling both MRRs to support both TE and TM modes. The calculated TE and TM mode profiles of the MRRs are shown in Figs. 2(b–c, e–f). The difference between the effective indices of the TE and TM modes (49GHz MRR: 1.627 for the TE mode and 1.624 for the TM mode, 200GHz: 1.643 for the TE mode and 1.642 for the TM mode) resulted in slightly different free spectral ranges (FSRs) for the TE and TM resonances and a wide spectral interval between them in the optical communications band (TE/TM separations of ~16.8 GHz and ~41.2 GHz for the 49GHz and 200GHz MRRs, respectively). The radius of the 49GHz MRR was ~592 μm, corresponding to an FSR of ~0.4 nm [34, 35] with a Q factor of $1.5 \times 10^6$, while the 200GHz MRR had a radius of ~135 μm, corresponding to an FSR of ~1.6 nm [36-39] and a Q factor of $1.2 \times 10^6$, which reduced leakage of the undesired sideband from the 200GHz MRR's unused resonances. The through-port insertion loss was ~1.5 dB after being packaged with fiber pigtails via butt coupling and employing on-chip mode converters. The devices have been shown to exhibit negligible nonlinear loss (two photon absorption) up to 25GW/cm2 [40]. The measured transmission spectra of the two MRRs are shown in Fig. 3, in which the FSRs and operation RF frequency are clearly reflected.

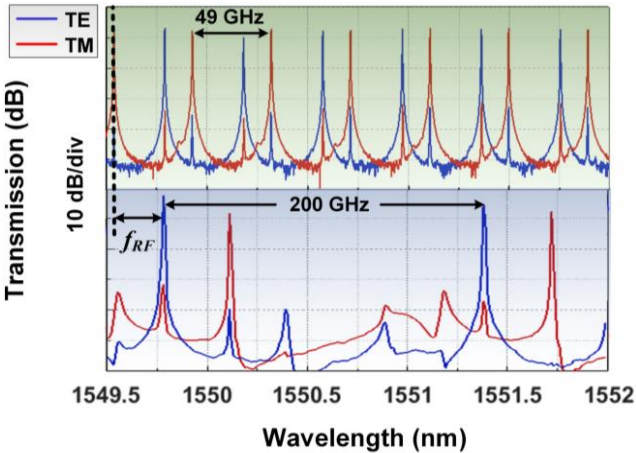

Fig. 3. Measured transmission spectra of the (49GHz FSR MRR) and 200GHz FSR MRR.

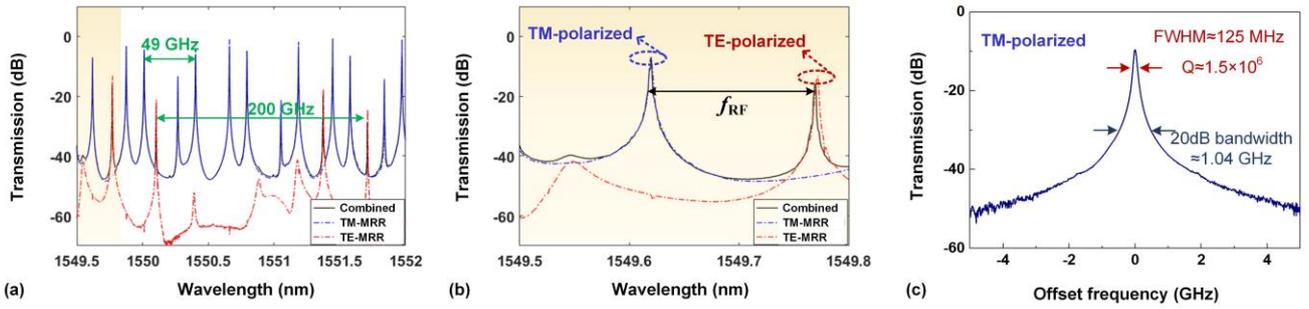

Fig. 4. Measured (a) transmission spectra of the 49GHz (TM) MRR, 200GHz FSR (TE) MRR, and the combined OP-OSSB generator. (b) Zoom-in spectra of (a) with one TE polarized resonance and one TM polarized resonance. (c) Transmission spectra around one TM-polarized resonance of the 49GHz FSR MRR.

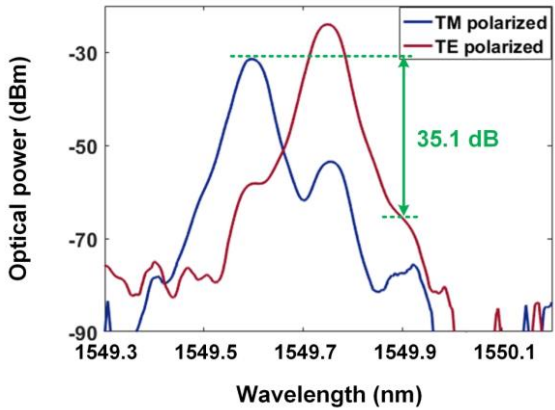

Fig. 5. Optical spectra of the generated orthogonally polarized OSSB signal.

The two MRRs were then connected via polarization maintaining fiber pigtails with the 49GHz MRR's through-port being connected to the 200GHz MRR's input. Both of the MRRs' drop-ports were then combined by connecting the drop-port of the 200GHz (TE) MRR to the add-port of the 49GHz (TM) MRR. Figure 4 shows the measured transmission spectra of the dual MRRs. As reflected by the dual resonances, both of the MRRs supported dual polarization modes. The first (49 GHz FSR) and second MRR (200GHz FSR) served as TM- and TE-MRRs for the OP-OSSB generation, respectively. The operation RF frequency was determined by the spectral interval between adjacent orthogonally polarized resonances (Fig. 4(b)). The 49GHz MRR featured a high Q factor, corresponding to a 20dB-bandwidth of ~1.04 GHz (Fig. 4(b)), indicating a high rejection ratio of the optical carrier and lower accessible RF frequency down to the sub-GHz level for the OP-OSSB generator.

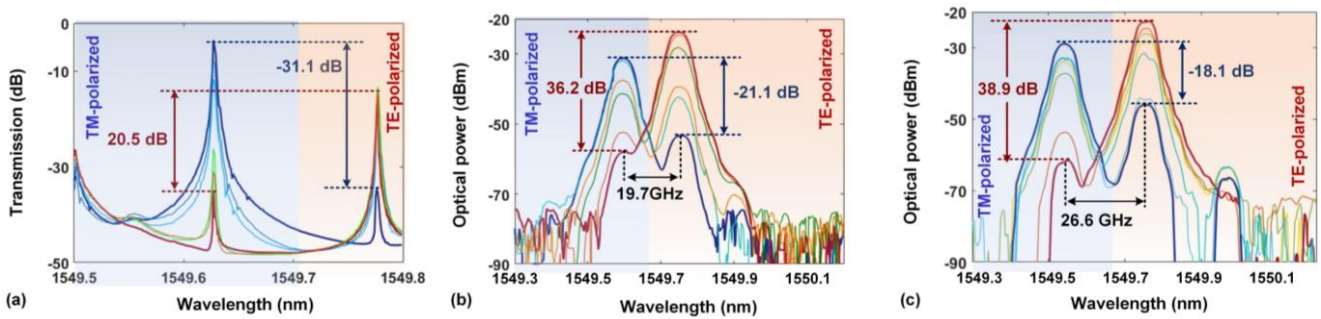

Fig. 6. Measured (a) transmission spectra of the dual MRRs and (b–c) optical spectra of the generated orthogonally polarized OSSB signal with continuously tunable OCSR.

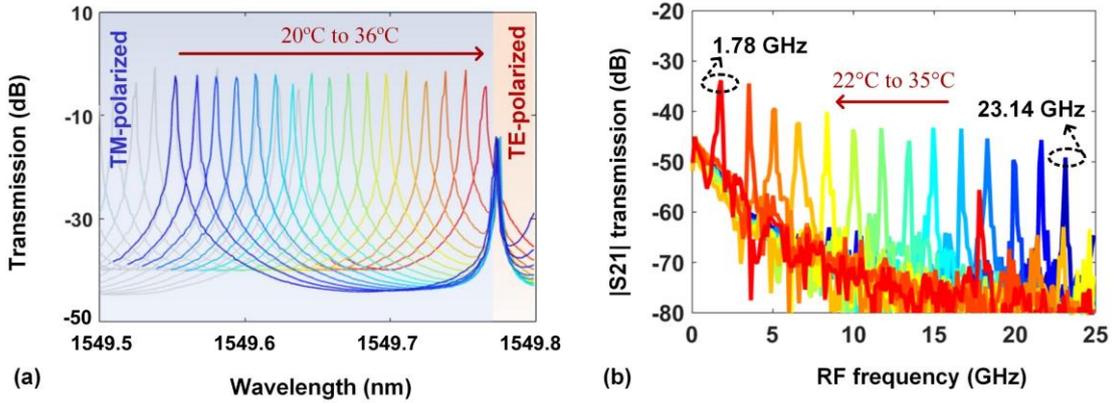

Fig. 7. Measured (a) optical transmission spectra, and (b) RF transmission response of the OP-OSSB generator with thermo-optical control.

During the experiment, we tuned the carrier wavelength to one of the TE resonances of the 200GHz MRR centered at ~1549.78 nm, and drove an intensity modulator (ixBlue) with an RF signal such that the lower sideband could be dropped by the adjacent TM resonance of the 49GHz MRR. The orthogonally polarized carrier and lower sideband were obtained at the output of the dual MRRs, where the optical power of the upper sideband was suppressed by over 35 dB as compared with that of the lower sideband, as shown in Fig. 5.

As mentioned, the ratio between the orthogonally polarized optical carrier and lower sideband could be tuned via the input polarization angle ($\theta$ in Fig. 1). We measured the corresponding transmission of the dual MRRs (Fig. 6(a)) as $\theta$ was varied. The extinction ratio between TE and TM resonances varied from 20.5 dB to –31.1 dB, corresponding to a dynamic tuning range of up to 51.1 dB for the OSCR. Figures 6(b)–(c) show the optical spectra of the generated OP-OSSB signal with input RF frequencies at 19.7 GHz and 26.6 GHz. Continuously variable OCSRs ranging from −21.1 to 36.2 dB and from −18.1 to 38.9 dB were achieved for the 19.7 GHz- and 26.6 GHz- RF input, respectively, yielding a large OCSR tuning range of up to 57.3 dB for our OP-OSSB generator. We note that the orthogonal polarization modes of the cascaded MRRs could also potentially provide an additional control dimension for optical logic gates, thus serving as a promising candidate for optical computing functions [41,42].

To achieve wide RF tunability, we varied the spectral interval between the TE resonances of the 200GHz MRR and the TM resonances of the 49GHz MRR by applying separate thermal controls to two devices [43]. The temperature of the 49GHz (TM) MRR was varied from 20 ºC to 36 ºC while the temperature of the 200GHz (TE) MRR was maintained at 25 ºC. Figure 7(a) shows the measured transmission spectra of the dual MRRs as a function of temperature, where the 49GHz MRR TM polarized resonance was thermally tuned over a range of 0.2 nm while the 200GHz MRR TE-polarized resonance was fixed, thus leading to over 20 GHz RF tuning range for our OP-OSSB generator. To reflect the wide RF operation of our approach, the OP-OSSB signal was converted into single polarization via a polarizer and detected by a photodetector. The RF transmission response of the system was measured by a vector network analyzer. As shown in Fig. 7(b), wideband RF operation up to 23.14 GHz was demonstrated. The optical spectra of the OP-OSSB signals with tunable RF operation were measured and shown in Fig. 8(a). As the chip temperature of the 49GHz (TM) MRR was varied from 22 ºC to 35 ºC, the RF operation frequency of the OP-OSSB generator changed from 1.81 GHz to 23.27 GHz with a fit slope of −1.66 GHz / ºC (Fig. 8(b)), thus confirming the wide tuning range of our approach. The RF operational bandwidth of our OP-OSSB generator was limited by the thermal controller (~15 ºC tuning range). Covering the entire FSR of the TM-MRR (49 GHz) would only require a thermal tuning range of 29.5 ºC which is readily achievable [43]. We note that the small change in FSR with temperature can easily be compensated for by the calibration of the device. Furthermore, by using multiply FSR spaced TM resonances, the RF tuning range can be increased arbitrarily, reaching even the THz region — well beyond that of electrical approaches [44].

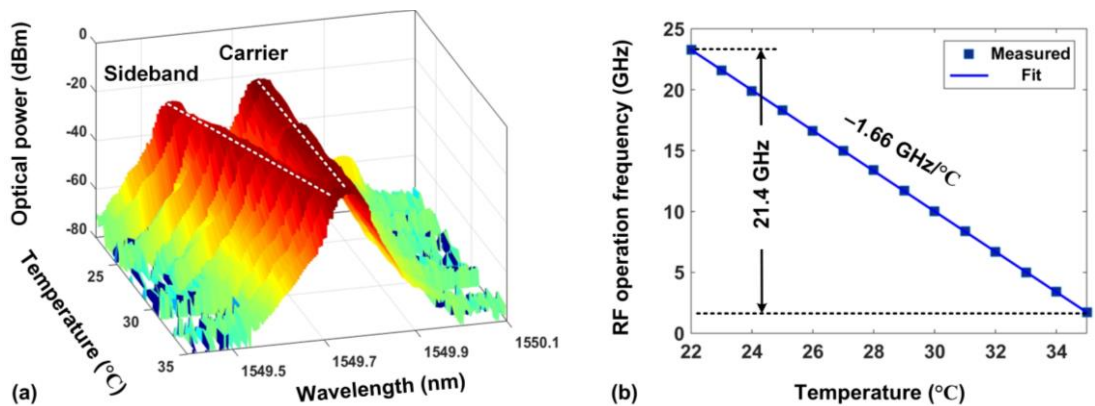

Fig. 8. (a) Optical spectra and (b) extracted operation RF frequency of the generated OP-OSSB signals with thermal tuning.

Since the cascaded micro-ring resonators are passive filtering devices, they did not contribute to the performance of the generated orthogonally polarized optical single sideband signals in terms of coherence/dephasing time. The dephasing time of the generated signals in our case was mainly determined by the coherence length $L_{coh}$ of our laser, given by [45]

$$L_{coh} = \sqrt{\frac{2\ln 2}{\pi n}} \frac{\lambda^2}{\Delta \lambda} \qquad (3.1)$$

where $\lambda$ is the central wavelength of the source (~1550 nm), $n$ is the refractive index of fiber (~1.45), and $\Delta\lambda$ is the full width at half maximum (FWHM) spectral width of the source. Our laser (TUNICS T100S-HP) had a 400 kHz FWHM spectral width, yielding a coherence length of ~414 m.

Here, we employed high-Q MRRs for the OP-OSSB generation, which support a narrow instantaneous RF bandwidth. For RF applications requiring a broad instantaneous bandwidth, either a lower Q MRR [46] or a higher order filter [47-50] can be employed to replace the high-Q TM-MRR in our experiment. The former can yield a 3dB bandwidth from 2 to 12 GHz corresponding to a Q factor ranging from 60,000 to 10,000, while the latter can achieve a 3dB bandwidth of 100 GHz or even higher. It should also be noted that the two MRRs can be further integrated on the same chip, with the spectral interval tuned by employing separate thermo-optical micro-heaters [51]. Finally, in this paper the micro-ring resonators based on this CMOS compatible platform are used as passive linear devices and so the issue of two photon absorption, important for nonlinear functions [52,53], is not relevant and so in principle any integrated optical platform with the appropriate linear optical properties could be used.

## 4. Conclusion

We propose and experimentally demonstrate an orthogonally polarized optical single sideband (OP-OSSB) generator based on dual integrated MRRs. By splitting the input double sideband signal into an orthogonally polarized carrier and lower sideband via TE- and TM-polarized MRRs, an OP-OSSB signal can be generated. A large tuning range for the optical carrier to sideband ratio of up to 57.3 dB was achieved by adjusting the polarization angle of the input light. The operational radio frequency of the OP-OSSB generator could be widely varied via separate thermo-optical control of the two MRRs, resulting in a broad RF tuning range of over 21.4 GHz. This approach provides a new way to realize OP-OSSB generation with wideband tunable RF operation, which is promising for RF photonic signal processing in radar and communication systems.

**Acknowledgments**

This work was supported by the Australian Research Council Discovery Projects Program (No. DP150104327). RM acknowledges support by the Natural Sciences and Engineering Research Council of Canada (NSERC) through the Strategic, Discovery and Acceleration Grants Schemes, by the MESI PSR-SIIRI Initiative in Quebec, and by the Canada Research Chair Program. He also acknowledges additional support by the Government of the Russian Federation through the ITMO Fellowship and Professorship Program (grant 074-U 01) and by the 1000 Talents Sichuan Program in China. Brent E. Little was supported by the Strategic Priority Research Program of the Chinese Academy of Sciences, Grant No. XDB24030000.